\documentclass{IEEEtran}
\usepackage{url,hyperref,lineno,microtype,subcaption}
\usepackage[onehalfspacing]{setspace}
\usepackage{comment}
\usepackage{graphicx} 

\begin{document}

\title{Revisiting the Internet of Things: New Trends, Opportunities and Grand Challenges}

\author{
    \IEEEauthorblockN{Khalid Elgazzar\,$^{1}$}
    , \IEEEauthorblockN{Haytham Khalil\,$^{1}$}
    , \IEEEauthorblockN{Taghreed Alghamdi\,$^{1}$}
    , \IEEEauthorblockN{Ahmed Badr\,$^{1}$}
    , \IEEEauthorblockN{Ghadeer Abdelkader\,$^{1}$}
    , \IEEEauthorblockN{Abdelrahman Elewah\,$^{1}$}
    , \IEEEauthorblockN{Rajkumar Buyya\,$^{2}$}
    \\
    \IEEEauthorblockA{\textit{$^{1}$IoT Research Laboratory, Ontario Tech University, ECSE, Oshawa, ON L1G 0C5, Canada}}\\
    \IEEEauthorblockA{\textit{$^{2}$Cloud Computing and Distributed Systems (CLOUDS) Laboratory, School of Computing and Information Systems, University of
Melbourne, Australia}}\\
    \IEEEauthorblockA{Khalid.Elgazzar@ontariotechu.ca}
}
\maketitle

\begin{abstract}
   The Internet of Things (IoT) has brought the dream of ubiquitous data access from physical environments into reality. IoT embeds sensors and actuators in physical objects so that they can communicate and exchange data between themselves to improve efficiency along with enabling real-time intelligent services and offering better quality of life to people. The number of deployed IoT devices has rapidly grown in the past five years in a way that makes IoT the most disruptive technology in recent history. In this paper, we reevaluate the position of IoT in our life and provide deep insights on its enabling technologies, applications, rising trends and grand challenges. The paper also highlights the role of artificial intelligence to make IoT the top transformative technology that has been ever developed in human history. 
\end{abstract}


\section{Introduction}

The Internet of Things (IoT) is a conceptual paradigm that connects billions of Internet-enabled devices to exchange data among themselves and their surroundings to enable smart interactions and connect the physical infrastructure to digital systems. IoT represents a revolutionary paradigm that started to affect our lives in many positive ways. The term Internet of Things was first coined in 1999 by Kevien Ashton \cite{Ashton} and was initially designed to support RFID technology. However, nowadays IoT has reached far beyond its designers’ vision and become much popular for the new applications it opens up in many vital domains like healthcare, intelligent transportation, public safety, home automation, smart city, asset monitoring, industrial automation and much more.  The evolution of IoT presented the long-awaited promise of ubiquitous data access in which people wanted to have access to real-time data on the go anywhere and anytime.

\par
Even though there are many other relevant paradigms/model that intersect with the purpose of IoT (e.g., M2M: Machine to Machine), Web of Things, Internet of Everything (IoE), pervasive computing, etc.), there are fundamental differences between them and IoT. The core values of IoT lies in the promise of helping businesses to increase their productivity, enhance control over their assets, and make informed business decisions based on the inference resulting from the processing of the fusion of big raw data acquired from the surroundings, including people themselves. Recent research statistics reveal over 10 billion connected IoT devices in 2021. This number is anticipated to reach 41 billion in 2027, expecting over 152,000 IoT devices to connect to the Internet per minute in 2025. Considering the global IoT market size, there was a 22\% increase in the market size of IoT in 2021, hitting \$157.9 billion. Smart home devices are the dominant components of IoT.  The penetration rate of IoT varies concerning the application domain. For example, IoT analytics \cite{RN142} argues that industrial applications occupy 22\% of the global IoT projects, with transportation, energy, and healthcare occupying 15\%, 14\%, and 9\%, respectively.

\par
The main two types of devices that make up the most of IoT are: Sensors and Actuators. Sensors are physical devices that can sense/measure a certain phenomena and can communicate the sensed values to other parties (i.e., collect data and report internal states). A GPS and an ECG device are examples of IoT sensors. The constituent sensor nodes usually utilize small-scale embedded systems to achieve the cost-effectiveness criteria of IoT solutions, increasing their deployment in various domains. Sensor nodes often use 8-bit microcontrollers and inhibit small storage capacity, lowering their power sizing and allowing them to run for years on batteries. Coupled with the diversified networking protocols available to match the existing infrastructure or the operational conditions, this highly promotes the deployment of IoT solutions in different domains. Actuators are also physical devices that can affect a change on the physical environments (i.e., take actions) in response to a command or a recommendation such as an AC thermostat and a valve. These devices need to be connected to the Internet and are able to communicate to send or receive data so they can qualify as IoT devices.

\par
The convergence of IoT, advanced data analytics and artificial intelligence opened up the door for the next generation of applications that support real-time decision making such as improved user experience and predictive maintenance. As such, data analytics has become a core component of any IoT deployment and will continue to gain popularity and relevance to businesses as much as data collection continues to grow and support intelligent decision making. In industrial manufacturing, for example, predictive maintenance can predict when maintenance is required in advance through the measurement of vibration levels, heat and other parameters to avoid production disruption. IoT data can also reveal rich information about customer behaviors (e.g., driving habits and shopping preferences) to support improved customer experience. Machine learning models and artificial intelligence techniques can learn from observations (IoT data collection) and recommend actions that lead into smart decisions (IoT actuation).    

\par
Although IoT promises to support intelligence decision making, enable better quality of life to citizens and make transformative changes in their daily lives, there remain grand challenges that hinder IoT from reaching its full potential such as privacy and security concerns, data heterogeneity and device interoperability, unrestricted access control and deployment in the open access domain.  The heterogeneity and small footprint of IoT of sensors for example, comes with two major shortcomings: 
(1) The constraints of resources available on the sensor nodes render it infeasible to apply the conventional security mechanisms typically involved in capable computer systems, exposing the sensor nodes as a weak security point for the whole IoT system. 
(2) The many networking protocols available to communicate sensed information among IoT devices result in interoperability issues between IoT systems utilizing different communication protocols.

\par
The first shortcoming of incapable sensor nodes results in the notorious ``vertical silos," where an IoT system is, in fact, a set of subsystems that lack information sharing among each other. That didn't represent a significant concern at the early ages of IoT since the applications were relatively limited, and the IoT had not reached its maturity and big vision yet. However, the advent of cloud computing in the last decade, coupled with the advancements in artificial intelligence and its subdomains, has vowed the prospect of IoT in various domains. This necessitates the ability of collaborative IoT systems to build better-informed business decisions based on the fusion of inferences coming from multiple systems. However, the second shortcoming of IoT impedes the scalability of IoT systems, confining the usability of sensed information by IoT systems to the managed networks of their users without exposing this information to public networks. This comes at the cost of increased IoT systems outlay, unwanted redundancy of the same information sensed by non-interoperable systems, expanded storage footprint, high network bandwidth utilization, risen processing cost, and more elevated system latency. This paper provides deep analytical views on many aspects of IoT technologies including standard architecture, stack protocols, value proposition, different IoT applications, trending technologies, and challenges. 

\par
The rest of the paper is structured as follows. Section \ref{IoT_Standard_Layered_Architecture_and_Protocols} discusses the IoT standard architecture, enabling technologies and stack protocols. Section \ref{IoT_Applications} describes the different domains of application for IoT with the most prevailing deployments. Section \ref{Rising_Trends_in_Sensor_Data_Analytics} sheds the light on rising trends in  IoT and the convergence between IoT and data analytics.  Section \ref{Open_Challenges} discusses the grand challenges for IoT that remain open for further research and deemed to decelerate its wide scale adoption. Lastly, Section \ref{Conclusion} offers concluding remarks.

\section{IoT Standard Layered Architecture and Protocols}
\label{IoT_Standard_Layered_Architecture_and_Protocols}

\begin{figure*}[!t]
\centering
\includegraphics [width = 7.0in] {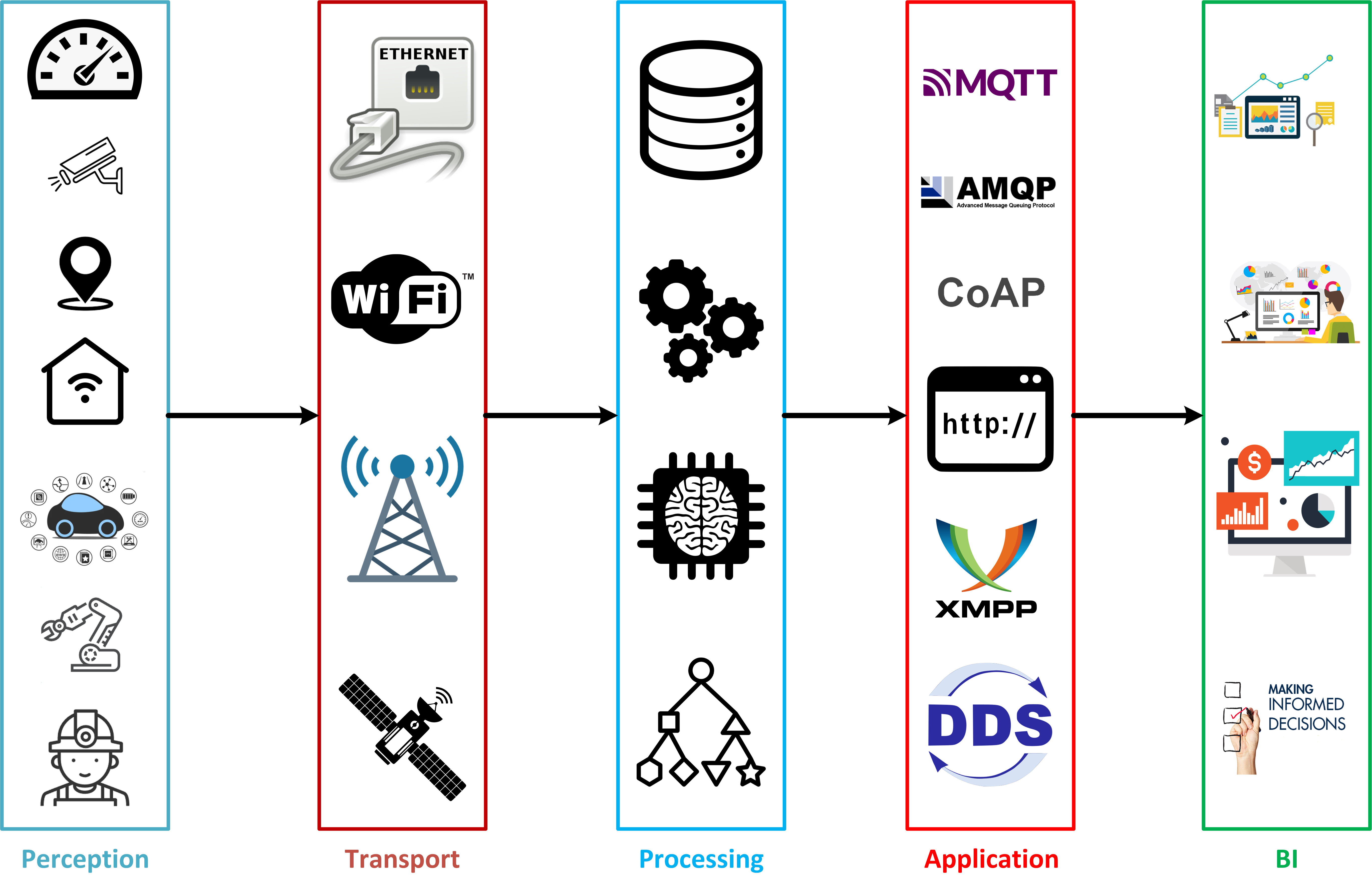}
\caption{IoT layered architecture.}
\label{figure:1}
\end{figure*}

From the engineering perspective, IoT is witnessing an increasing number of enabling technologies. This high diversity of IoT enabling technologies stem from the proliferation of IoT devices, their heterogeneity and uncertainty of operational environments, the advancements in chip manufacturing, and variety of communication protocols \cite{RN84}. Nonetheless, the advent of artificial intelligence (AI) and associated machine learning (ML) techniques leverage the serendipity of IoT by providing insightful information from the fusion of raw data collected by heterogeneous sensors to support decision making and change how people carry out their everyday business. This adds up to the enabling technologies of IoT. Therefore, abstracting IoT systems in terms of building blocks helps to contrast the hazy boundaries between different enabling technologies and enhance the agility and robustness to achieve a successful paradigm for IoT systems \cite{RN85}. The core elements of a typical layered IoT architecture, as depicted on Figure \ref{figure:1}, can be summarized as follows.

\subsection {Perception Layer}

The first layer of the IoT architecture is the perception layer, also denoted by the hardware, physical, or infrastructure layer. This layer encompasses the constituent physical devices of an IoT system that are typically responsible for: (i) sensing the environment in their vicinity and sending the raw sensed data to the next upper layer for processing, such as environmental sensors; (ii) transforming the logical decisions coming from the upper inference layer into physical actions applied to corresponding devices, such as actuators and servo motors. It is worth noting that, and as the name of IoT implies, the constituent devices that form an IoT system embed some form of communication by which they can be directly or indirectly, with the help of a gateway, connected to the Internet. Moreover, IoT devices typically include some form of identification that helps differentiate the data passed to the upper layers of the IoT architecture. This identification can be either burnt into the device firmware by the manufacturer (such as the unique identifier (UUID)), set up by the user through configurable menus or DIP switches, or provided by the communication subsystem that the devices utilize (like the MAC address or the Bluetooth identifier). 

\subsection {Transport Layer}

The transport layer, also denoted by the communication and network layer, and as its name implies, is responsible for connecting IoT devices in the perception layer to the upper layers of the IoT architecture, which are typically hosted over the Internet using cloud computing technologies. This layer utilizes a wide range of communication technologies, like cellular, Wi-Fi, Bluetooth, Zigbee, etc. Besides, the transport layer is responsible for maintaining the confidentiality of the data exchange between the perception layer and the upper layers. Nonetheless, with its potential promise and anticipated ubiquity and prevalence, IoT is the motivating force behind recent research in enabling communication technologies. For example, IPv6 has been identified such that it can provide network addresses to the anticipated enormous smart objects connected to the Internet, which exceeds the already depleted IPv4 addresses. Similarly, the 6LoWPAN communication standard has been mainly developed to enable IPv6 packet transmission for power-constrained smart objects communicating over IEEE 802.15.4.

\par
The transport layer securities typically used for IP-based networking, namely Transport Layer Security (TLS) and Datagram TLS (DTLS), provide the essential means for secure end-to-end communication. However, these technologies are not always feasible for deployment in resource-constrained embedded IoT devices due to the induced increased processing, storage, and power consumption overhead associated with these security mechanisms. This, in turn, usually delegates the authentication and the data integrity tasks of exchanged information in IoT systems to be arbitrarily carried out by the application layer based on the required security level and device capabilities. Besides, it exposes these poorly-secured IoT devices as a weak point for malicious users to penetrate the underlying critical network infrastructure or exploit them for botnet attacks to prevent the availability of network resources, aka distributed denial of service (DDoS). Derived by the proliferation of IoT devices, recent statistics anticipate that more than \%25 of all cyberattacks against businesses will be IoT-based by 2025. This slows down the adoption of IoT and makes businesses reluctant to expose the reachability of sensed information by their IoT systems beyond their managed networks, adding up to the ``isolated islands" dilemma of IoT systems.

\subsection {Processing Layer}

The processing layer, also denoted as the middleware layer, encompasses advanced features that could not be embedded within the inherently resource-constrained devices at the perception layer. This includes storage, processing, computing, and action-taking capabilities. Besides, the middleware layer facilitates IoT systems' scalability and interoperability across the computing continuum from the edge to a remote cloud data centre . It typically provides interfaces, like APIs, for other systems and third-party services to leverage the gathered raw data from the IoT devices or the insight obtained by the middleware layer after data processing. Based on the agreed tradeoff between device loads and bandwidth during the system design phase, the middleware layer can be either embedded within an on-site capable embedded platform, sometimes denoted as an IoT gateway, or hosted over the cloud. The former requires utilizing a medium-to-large scale embedded device to act as a gateway. Nonetheless, it typically utilizes a Linux kernel-based OS to mask the complexity of the underlying hardware interfacing to the perception layer devices. The latter, however, depends on relaying the raw data from the perception layer to cloud-hosted servers. This comes at the cost of higher bandwidth utilization and increased latency.

\par
The emergence of a cloud-hosted middleware layer for IoT systems represents a bottleneck considering the security concerns of IoT. Cloud computing is the only candidate to digest the enormous amount of IoT data coming from perception layer devices. However, cloud providers are also principal targets for cyberattacks and single points of failure for IoT systems. A successful cyberattack could expose an enormous amount of sensitive information to hackers and render the IoT system unfunctional. This puts system designers in a tradeoff of choosing between the capability, cost-effectiveness, and ease of access associated with cloud computing technologies on one hand and the panic of potential data leakage in case of a successful cloud attack on the other hand.

\subsection {Application Layer}

The application layer defines the domains by which IoT systems are deployed. This includes smart homes, smart cities, smart agriculture, etc. The application layer manages the logical processes to be taken based on the inference coming from the middleware layer and the system requirements. This includes sending emails, activating alarms, turning a device on or off, setting parameters, etc. Therefore, the application layer represents the user interface to interact with the other layers below it, facilitating human-machine interactions. Since the application layer is meant to be used by people, it inhibits a wide surface area exposed to good actors and bad actors. The common vulnerabilities usually encountered in the application layer include distributed denial of service (DDoS), HTTP floods, SQL injections, and cross-site scripting. Although large-scale cybersecurity attacks are dangerous, the effect of small-scale cybersecurity attacks, usually encountered in IoT systems, can be even more dangerous. This is because they do not have unique ecosystems, their cyber defense has not yet reached maturity, and they can be gone unnoticed for a long time. The security mechanisms applied at the application layer are meant to fulfill the CIA triad, namely confidentiality, integrity, and availability. This implies keeping the secrecy of exchanged information between communicating parties, ensuring that no alterations have been maliciously carried out on the information from its source to destination, and making the information always available to authorized users requiring it.

\par
In contrast to the vulnerabilities in the lower layers of the IoT architecture mentioned above which can also affect the upper layers, security breaches in the application layer do not affect the lower layers. In capable computer systems, however, security mechanisms are applied in parallel to different layers to tighten the system's safety. Nevertheless, for a market usually biased towards the price and the convenience rather than the security, this is not usually valid for constrained embedded devices often encountered in IoT systems. Security practices for IoT systems often delegate the security measures to be applied at the application layer based on the system requirements or even delegated to the third-party firewall appliances managing the network. Security measures in IoT systems usually come at trade-offs regarding the capacity of the constituent IoT devices utilized by the system. Moreover, with the diverse application domains of IoT, security mechanisms can even affect the system's effectiveness. For example, a VoIP-based IoT solution can be adversely affected by the induced latency of the security mechanism in action. On the other hand, this latency pales for confidentiality- and integrity-critical applications, like financial and medical applications, where the effect of a security breach would be catastrophic. 

\subsection {Business Intelligence Layer}

A successful IoT system depends on the utilized enabling technologies and how inference is delivered to the user abstractly and efficiently. The business intelligence layer is meant to fulfill this task by providing the user with visualized representations of the information coming from the middleware layer, masking its complexity and making it easier for the user to make informed business decisions.

\par
The business intelligence layer is not affected by the constituent embedded IoT devices utilized by the system. It does not deal directly with the constituent IoT devices. However, it deals with the inference from the middleware layer after processing the raw data from the IoT devices through the application layer protocols. Therefore, the security of the business intelligence layer depends on the typical user-level security mechanisms found in capable computer systems. The user-level security can be applied to different entities constructing the layer. This includes files, databases, or any other resources. The user-level security is meant to implement a fine-grained authorization control over accessible information to different users based on their credibility. 

\section{IoT Applications}
\label{IoT_Applications}

\begin{figure*}[!t]
\centering
\includegraphics [width = 7.0in] {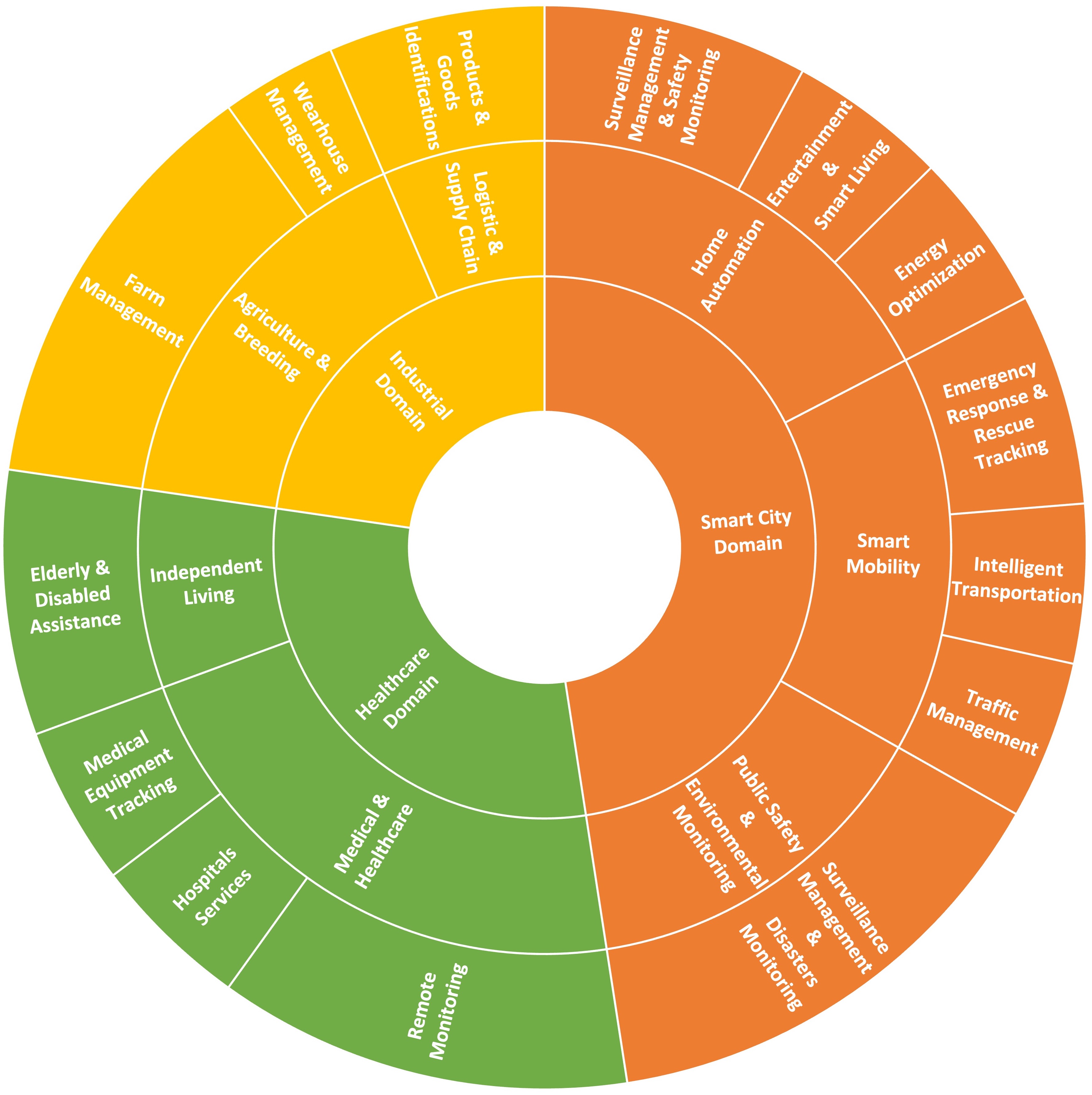}
\caption{IoT application domains and related services.}
\label{figure:2}
\end{figure*}

IoT can be seen in different real-world applications and services such as home automation, intelligent transportation, smart cities, digital healthcare, remote health monitoring, smart agriculture, and industrial automation \cite{RN150}. In each application domain, several sensors are triggered to independently gather data, transmit information, and initiate and execute services with minimum human intervention \cite{RN88}. The main objective of integrating IoT technology into real-world applications is to enhance the quality of life. For example, in the domain of smart city services, we find IoT applications used for increasing city safety, efficient mobility, and enhancing smart energy usage \cite{RN89}. On the other hand, IoT technology has introduced remote medical monitoring systems in the healthcare domain which empower physicians to provide superior care to patients \cite{RN90}. Numerous research proposes different IoT-based solutions and innovations under three application domains shown in Figure \ref{figure:2} \cite{RN88}.

\subsection{Smart City}

IoT technology assists cities to enhance mobility services, improve public safety, and control and automate household systems. Intelligent transportation, for example, focuses on solutions that manage road infrastructure and improve route planning for drivers. Furthermore, it provides innovative solutions to monitor and manage traffic systems using smart traffic signals and sensors, throughout the road network to smooth the traffic flow and reduce congestion. The concept of smart city services is not restricted to transportation, but also involves other aspects of human life, such as public safety, green and clean environment, smart grid, efficient delivery of municipal services and connecting the physical infrastructure to the digital world. Recent attention has been particularly paid to the fields of traffic management, intelligent transportation, and emergency response. A more detailed description of the use of IoT technology in these three fields is given in the following subsections.

\subsubsection{Traffic Management}

IoT-based traffic management systems mainly monitor road traffic conditions to solve the problem of increased traffic congestion and predict traffic status \cite{RN91}. These systems assist drivers by informing them about the traffic conditions at a given location and time. For example, an adaptive traffic signal control (ATSC) system that captures the traffic volume level can significantly reduce traffic congestion \cite{RN92}. Equipping roads with advanced sensors that capture real-time traffic data assists in determining the duration of traffic light signals across intersections. The ATSC system not only eases the traffic flow at intersections but also reduces travel time and fuel consumption, contributing positively to green environments. Research on IoT-based real-time ATSC systems at an intersection describes the coordinated approach as it is used to track the movements of vehicles and pedestrians \cite{RN95, RN93}. It also uses the deep reinforcement learning (DRL) method which is commonly used in ATSC systems to teach/educate traffic controllers how to make proper decisions. DRL simulates the effect of a traffic signal's action and the resulting changes in traffic status. It would be classified as an appropriate action in that given situation if the action improved traffic conditions, and as a negative action otherwise \cite{RN93}.  Furthermore,  IoT-based traffic management systems can facilitate smart parking in public spaces such as on-street parking, and lot parking. According to Libelium company \cite{RN300}, parking spaces play critical roles in reducing traffic volume, and gas emissions. In smart parking, drivers can easily locate an empty parking place using smart parking maps. These smart maps use IoT sensors and cameras to detect and manage the likelihood of parking space in a given area. An example of a real-time traffic occupancy system for smart parking is called SplitParking which is managed by the city of Split in Croatia \cite{RN89}. The SplitParking system places sensors integrated with an IoT technology within its parking spaces to monitor space occupancy andlert the end user of the parking availability through a user-friendly mobile application \cite{RN89}.

\subsubsection{Intelligent Transportation}

The emergence of IoT has provided a new perspective for intelligent traffic systems development. This is because the IoT paradigm satisfies the public’s demand towards an ‘’always connected’’ model by relying on the interconnection of our daily physical objects using the Internet.  Hence, allowing it to collect, process and transfer data creating smart intelligent systems without human intervention.

\par
In the traffic domain, IoT requires every element such as roads, tunnels, bridges, traffic lights, vehicles and roadside infrastructures to be Internet-connected for identification and management purposes. This can be done using sensor-enabled devices, for instance, RFID devices, Infrared sensors, GPS and many others.  Intelligent traffic systems that are IoT-based can efficiently improve traffic conditions, reduce traffic congestion and are unaffected by weather conditions. Moreover, IoT allows for dynamic real-time interactions, since it facilitates the incorporation of communication, control and data processing across the transportation systems. Beyond any doubt, IoT is causing a noticeable shift in the transportation sector.
 
\par
The rapid advancements within information and communication technologies have also paved the way for developing more self-reliable and intelligent transportation systems. These include striding advancements in hardware, software, sensor-enabled and wireless communications technologies.  Therefore, moving towards a new era of connected intelligent transportation systems, where the demand for on-going and future real-time traffic data continues to rise \cite{RN99}. This enforces several challenging requirements on the traffic information systems. Among these requirements is broadcasting real-time, user-friendly, and precise traffic data for users. These traffic data including color-coded maps showing congestion, calculated traffic time intervals between arbitrary points on the road network. In addition to traffic density estimations that should be easily interpreted by users in a very short time. Moreover, demonstrating real-time routes for drivers based on the embedded navigation systems such as Global Positioning Systems (GPS). Another challenging requirement lies in storing huge amounts of traffic information generated from progressively complex networks of sensors, where Big data comes into play. However, collecting and storing this amount of traffic  information is not enough.  It is crucial to correlate based on game theory methodology, validate and make use of data in real-time. Hence providing valuable, relevant data extraction and insightful predictions of upcoming patterns and trends based on historical data. As an example, providing drivers with real-time traffic information to assist them in finding out the best road routes. This is why the subsequent role of predictive analytics for the whole transportation is needed. Consequently, traffic information systems require participation of all of the above components to interact and integrate through a common infrastructure. It allows immediate transmission of real time traffic information to any part of the system \cite{RN301}.

\begin{figure*}[!t]
\centering
\includegraphics [width = 7.0in] {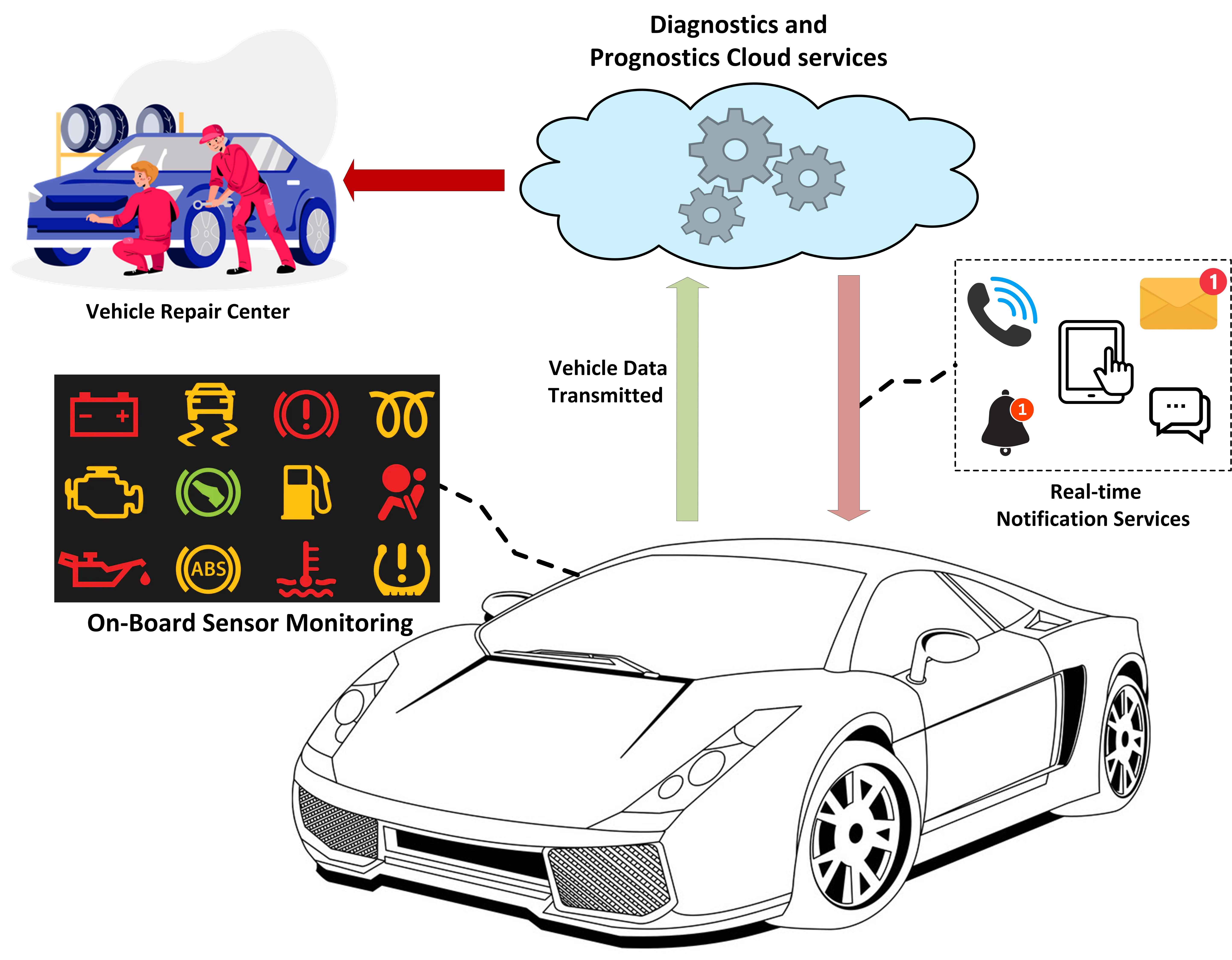}
\caption{An IoT-based predictive maintenance use-case scenario.}
\label{figure:3}
\end{figure*}

\par
Even with the aforementioned benefits and challenges that come along with the integration of IoT into the transportation sector.  IoT provides a paradigm shift that changes the transit services into  intelligent groundbreaking systems, where numerous  cutting edge technologies are incorporated. This creates a wide suite of intelligent transportation applications that have road users’ experience and safety at its core. A widely adopted IoT applications in the automotive industry include: the integration of sensors such as weight measurements and real-time fleet location sensors-based tracking to help fleet operators efficiently manage their fleets. Another use case that IoT technologies have shown great impact is predictive analytics.  In this context,  drivers are provided with early in advance vehicle maintenance alerts in cases of failure of a specific vehicle component. This is because these components are equipped with sensors  that collect and share real-time information on the vehicle’s status with their vendors. It avoids any sudden or abrupt failure that can cause a life-hazard situation.  Last, but not least with a precise focus on the integration of IoT with connected mobility. Figure \ref{figure:3} showcases a predictive maintenance scenario, where an in-vehicle monitoring system acquires IoT sensing data from the faulty in-vehicle sensor. The vehicular data is sent to the Diagnostics and Prognostics cloud services for analyzing and predicting maintenance issues. Repair recommendations are then sent back to the drivers \cite{RN115}. 

\subsubsection{Emergency Response}

Crisis management is one of the critical situations that face many governments, first responders, emergency dispatchers and others who provide necessary first aid/assistance at the least possible time. In such situations, the design of the required infrastructure to handle emergencies becomes a critical requirement. With the introduction of IoT technologies to the safety systems where a suite of sensors are connected to provide real-time data to crisis management officials. This includes the use of sensors to monitor the water levels in cases of flood situations to provide insights that can support real-time data analytics to manage flooding crises. Real-time information can contribute to improving crisis management response time. Hence, reducing  or eliminating the costs of crisis-related damages. Firefighting is considered a viable use case that finds value in IoT applications. Heat-proof sensors placed in indoor buildings can provide real-time information about the initial starting point of a fire, spreading patterns and intensity levels \cite{RN116}. It also extends to provide additional safety measures for firefighters as they use IoT-based safety alert devices that can  accurately detect their motions. These devices are equipped with acoustic transmitters which act as beacons to allocate firefighters within the building and embedded sensors that can monitor their vital health conditions.  Besides protecting firefighters, IoT-based sensors are employed to sustain indoors electrical systems and smartly pre-identify any active heat sources through abrupt temperature spikes. Immediate alerts are then subsequently sent for rapid and instant inspections.  Fire systems based on IoT solutions can be actively intelligent to detect and promptly put off small fires through the use of smart sprinklers. Other emerging solutions that aid firefighters in fire crisis situations include computer-aided dispatch data such as precise fire locations, environmental conditions and others. Augmented reality-IoT based firefighter helmets \cite{RN302} are another innovative solution that can effectively guide firefighters to navigate in low-visibility conditions.

\par
Emergency responders and dispatchers are leveraging the benefits of IoT-based solutions when dealing with daily traffic accidents. Numerous automotive industrial solutions such as GM OnStar provide a myriad of applications and services to assist  first aid dispatchers. This could be achieved through utilizing cellular networks in conjunction with GPS and IoT technologies. Leveraging Vehicle to Infrastructure (V2I) communication-based technologies to provide critical information incases of traffic accidents. These include avoidance crash response, where drivers in crash situations can connect to OnStar call center by requesting the appropriate help to be provided to the vehicle's location. The ecall can be activated manually (using a push-button) or automatically through data collected from on-board sensors. Other applications include stolen vehicle assistance which help the authorities in locating the stolen vehicle by activating several functionalities. This  includes halting the restart option upon reactivation of the remote ignition block and transmitting a slowdown signal to let the vehicle come to a stop eventually \cite{RN121}. Other use cases include amber alert notifications sent to the public by integrating IoT and cellular network technologies.  Amber alerts provide new means of aiding emergency responders and authority officials in risky situations such as child abduction.  Officials collect crowdsourcing witness information from people within close event proximity to assist in their investigations. The alerts usually include event description (time and location). In addition to the  Kidnappers’ detailed information (e.g,  vehicle’s information, license plate number and their description) as well as child description. However, inability to correctly track the suspects’ vehicle or missed notifications by the public may contribute and lead to inefficient amber alert-based systems \cite{RN94}.

\par
Traditionally, infrastructure failures and power outages are other use cases implying sudden and abrupt crisis situations that may be disruptive to emergency officials.  Based on leveraging IoT technologies that aim at providing preventive and predictive maintenance. Hence, avoiding sudden breakdowns, anomalies and damages of the infrastructure. This could be achieved through continuous supervision and monitoring. For instance, smart bridges include a modular and IoT-sensor based system for monitoring, evaluating and recording any changes of the bridge structure in near real-time. Embedded sensors in the core structure of the bridge can then relay measurable information to management officials for further analysis . This includes humidity, temperature and corrosion status of the structure . Such data is considered valuable to constantly assess and evaluate the health structure and provide necessary measures such as intervention and predictive measures strategies \cite{RN117}. 

\subsection{Home Automation}

Home automation and control systems are essential components of smart cities and have played a significant part in the advancement of our home environments. They have several applications for different usage at home, such as entertainment and smart living, surveillance, and safety management \cite{RN97}. Home automation is described as a standard home environment equipped with IoT technological infrastructure to provide a safe and comfortable lifestyle \cite{RN98}. Home automation is based on an intelligent, self-adaptive system that analyzes and evaluates stakeholder behaviors and has the capability to predict the stakeholder's future actions and interact accordingly. Home automation systems use image detection and facial recognition models that are embedded in an intelligent control system connected to different sensors such as light sensors, motion sensors, water leak sensors, smoke sensors, and CCTV cameras \cite{RN100}. These devices communicate with each other through a gateway that is distributed throughout a home area network. The home control system will connect different subsystems that cooperate in modeling the stakeholder’s actions and the environment's information such as temperature, humidity, noise, visibility, and light intensity to enhance the learning process. For example, lights and AC temperature can be controlled and automated to adapt to the stakeholders' needs and their movements in the home environment. This would conserve energy while also effectively monitoring energy consumption \cite{RN104}. Research on home automation is not restricted to energy optimization; it involves health monitoring and security measures. By using innovative IoT technologies, we can connect to surveillance cameras in the home environment via a mobile device. Additionally, stakeholders can have access to doors and window sensors to maintain home safety and security remotely \cite{RN105}. 

\begin{figure*}[!t]
\centering
\includegraphics [width = 7.0in] {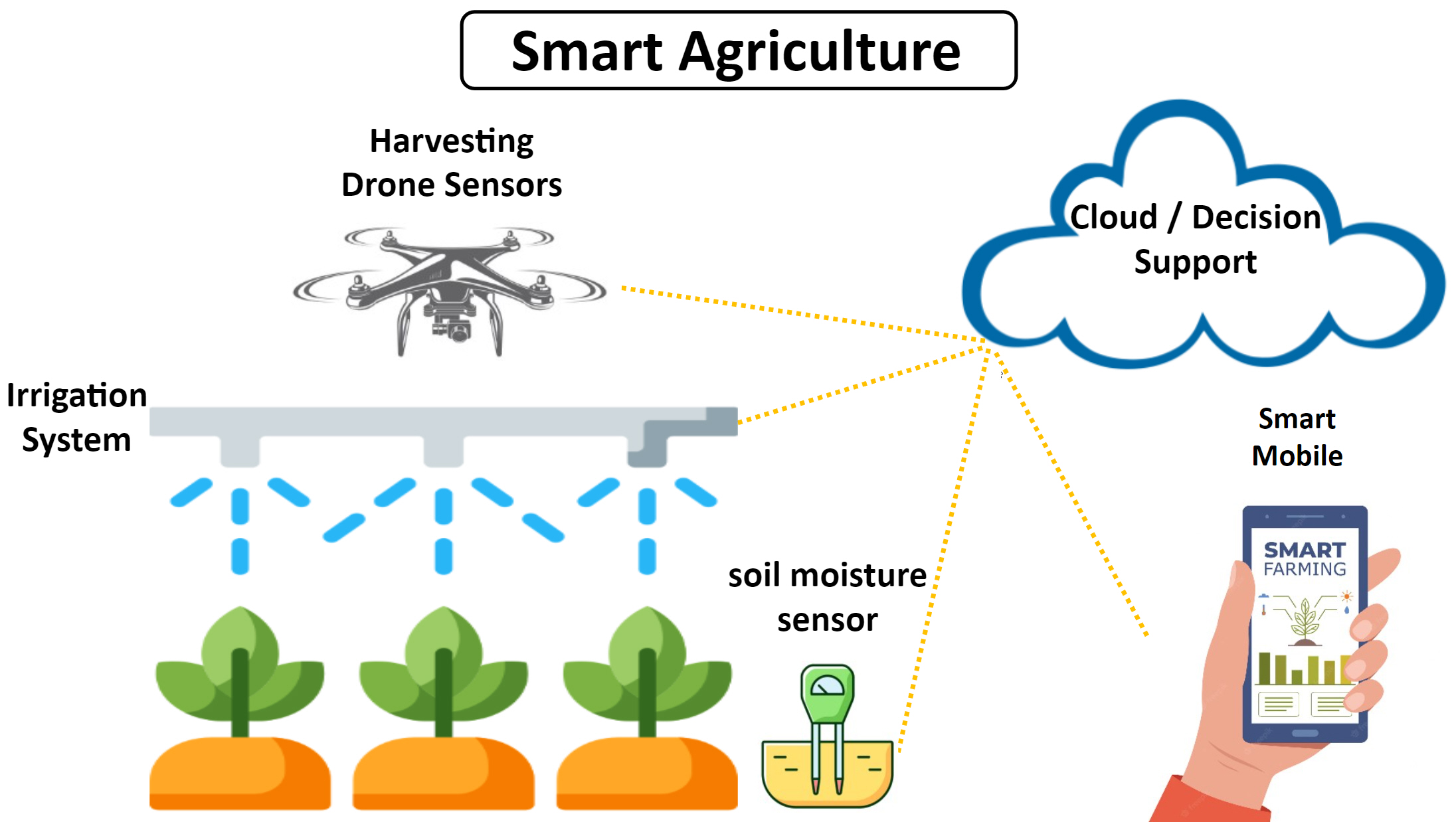}
\caption{Illustration of a smart agriculture system.}
\label{figure:4}
\end{figure*}

\subsection{Industrial Sector}

Industrial IoT leverages IoT capabilities in business and economic sectors to automate previously complex manual operations in order to satisfy consumer needs and reduce production costs. Warehouse operations, logistical services, supply chain management, and agricultural breeding can have machine-to-machine (M2M) intercommunication to ensure optimal industrial operations \cite{RN304}. Figure \ref{figure:4} illustrates a scenario of the IoT communication sensors in a smart agricultural system. This smart agriculture system monitors and analyzes the environmental parameters using soil moisture and harvesting sensors such as ZigBee, EnOcean, Z-wave and ANT \cite{RN96}. These sensors are automated to diagnose the status of a plant and gather this data through an IoT platform to take the proper action such as when to irrigate in consultation with a weather forecasting service available in the Cloud; thus ensuring the efficient use of water resources.

\begin{figure*}[!t]
\centering
\includegraphics [width = 7.0in] {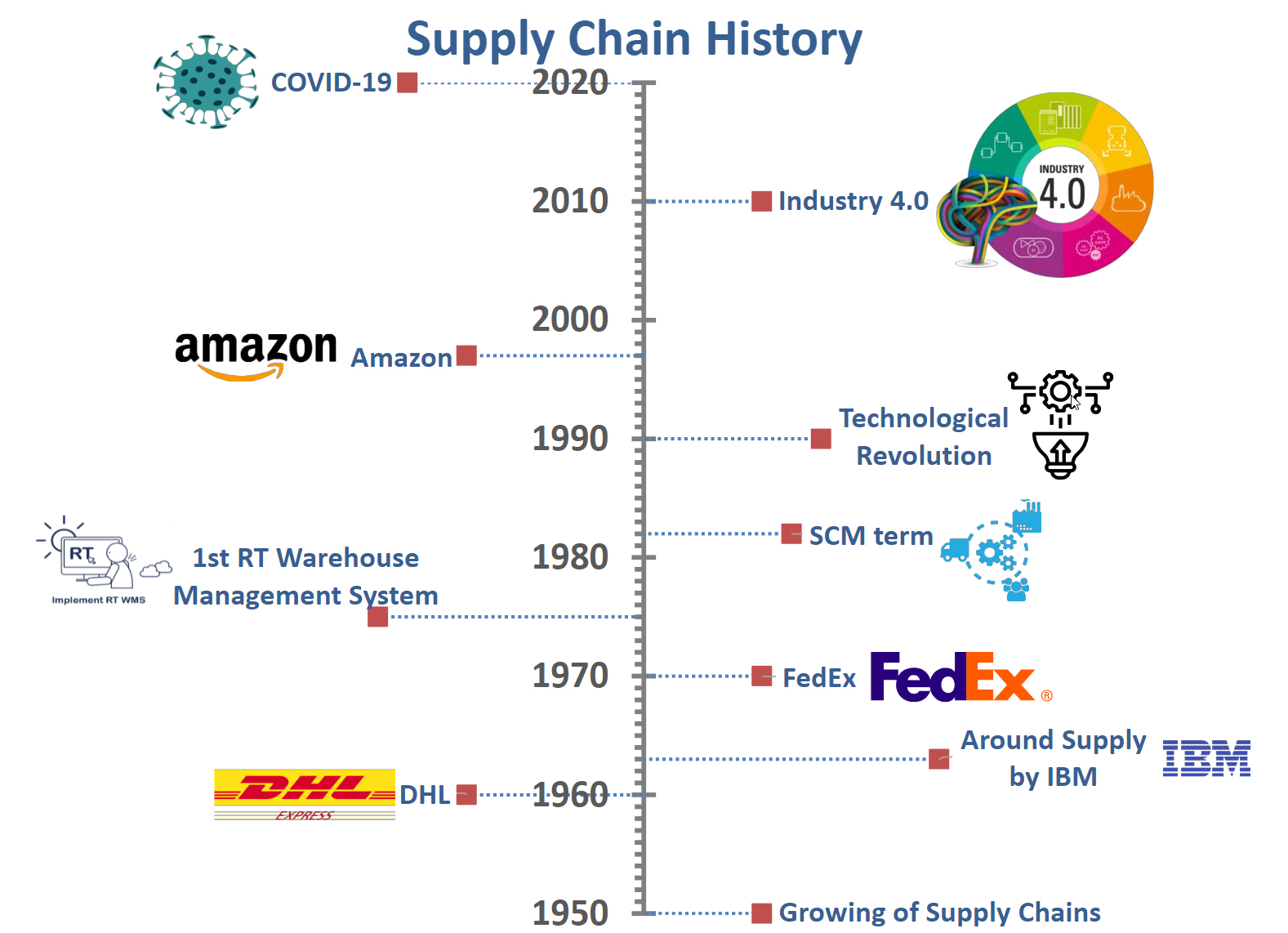}
\caption{Supply chain milestones.}
\label{figure:5}
\end{figure*}

\subsection{Logistics and Supply Chain}

Supply Chain Management (SCM) is a crucial service in our world. Since 1900 \cite{RN147}, humanity has evolved SCM to meet the market needs. Figure \ref{figure:5} highlights SCM milestones. Before 1900, SCM was restricted to the local areas. However, due to the revolution in railways, goods now can reach far beyond local borders. Between 1900 to 1950, global SCM attracted large players and organizations like UPS began providing their services in the SCM field. Industry leaders started to look for how we could improve the mechanization of the SCM process. From 1950 to 1970, the SCM community gained a superior experience by analyzing the military logistics of the First World War. DHL and FedEx were established as logistics enterprises, and IBM built the 1st computerized inventory management that was capable of handling complex inventory problems and making stock forecasts. In 1975, JCPenney designed the first Real-Time Warehouse Management System (WMS) that monitors the warehouse stock in real-time. Seven years later, Keith Oliver introduced the Supply Chain Management term. In the 90s, the technological revolution  was triggered  when many enterprises deployed computers to manage their processes and the internet to reach their customers through the World Wide Web. In the 90's, Amazon started running the e-commerce website. The 4.0 industrial revolution, including the internet of things, began growing in the last decade. Although the IoT looks like a promising technology to be adopted in the SCM field, deploying IoT in SCM faces many challenges. The main hindrance \cite{RN144} is the integration of different supply chain processes due to The heterogeneity of technologies used in various supply chain stages.

\par
COVID-19 \cite{RN146} uncovered a new factor that magnifies the importance of relying on information and communications technology to run the SCM systems. Businesses had to switch to remote working due to the pandemic. The lockdown and physical distancing requirements imposed on suppliers reduced their labors in their plants and sometimes obliged to shut down to limit the virus spreading. As a result, consumers face product shortages due to reduced production volumes during the pandemic. To date \cite{RN145}, the world is still suffering from the devastating effects of COVID-19 on the supply chain. Therefore, decision-makers \cite{RN148, RN149} started exploring how we could deploy new technologies, such as IoT, for managing remote operations.

\subsection{Healthcare Sector}

IoT sensors and devices shifted the landscape of portable and wearable medical devices from fitness and wellness devices to medical-grade devices qualified for usage at hospitals and healthcare providers. This shift accelerated the integration of remote patient monitoring in hospitals to accommodate patients with chronic diseases \cite{RN108}. Therefore, numerous efforts have been conducted to advance remote patient monitoring (RPM) systems with the help of well-established IoT infrastructures and standards in the healthcare domain \cite{RN136}. The RPM systems are expected to match or exceed the performance of the existing monitoring and examinations administered at hospitals and healthcare facilities \cite{RN108}. For example, continuous heart rate monitoring and immediate heartbeat detection necessitate patients to be hospitalized and/or connected to a Holter monitor or similar devices for long-term cardiac diagnosis. However, this setup would hinder patient mobility due to the limitations of the existing devices in terms of size and the number of attached wires. Moreover, hospitals dedicate significant resources to providing long-term cardiac monitoring that, in some cases, is unavailable, especially in low or middle-income countries. Therefore, RPM systems effectively reduce death from chronic diseases (e.g., heart diseases, diabetes). IoT platforms and devices significantly accelerated the development and integration of RPM systems into existing healthcare infrastructures. To that extent, a typical RPM implementation constitutes various services but is not limited to data acquisition, tracking, communication, automated analysis, diagnoses, and notification systems \cite{RN137}.

\section{Rising Trends in Sensor Data Analytics}
\label{Rising_Trends_in_Sensor_Data_Analytics}

In recent years, the IoT domain has witnessed increasing interest by the research community and rising demands from the industrial sector to embed real-time data analytics tools into the core of IoT standards. While the real value proposition of IoT is shifting from providing passive data monitoring and acquisition services to autonomous IoT applications with real-time decision-making services. Consequently, real-time data analytics is no longer an add-on service and has become integral to any IoT application rollout. For example, remote patient monitoring (RPM) and real-time data analytics have significantly contributed to enhancing ECG monitoring and enabling healthcare providers to gain 24/7 access to their patients remotely, especially for patients with coronary diseases \cite{RN134}. However, sensor data acquisition and collections are mapped as the foundation of IoT applications yet are considered passive techniques due to the absence of intelligence or decision-making. The main goals of IoT application at the early stages were to collect and monitor significant information regarding specific applications as initially proposed in 1999 \cite{RN133} while developing supply chain optimization at Procter \& Gamble. Nearly after two decades, the goals of using IoT applications and their expectations are on the rise, demanding proactive and active decisions made on sensor data collected in real-time. Accordingly, data analytics permits various applications to focus on performing real-time diagnoses, predictive maintenance, automated decision-making, and theoretically improving the productivity and efficiency of the intended applications. Meanwhile, modern stream processing engines (e.g., Apache Kafka and Apache Pulsar) come with built-in APIs ready for data analytics integrations \cite{RN135}. Moreover, most cloud services provide ready-made end-to-end event processing and real-time data analytics tools (i.e., Google DataFlow).

\subsection{Real-time vs Offline Data Analytics: Differences, Needs, and Potential Use}

In an IoT-driven society, applications and services integrate smart learning approaches for analyzing insightful patterns and trends that result in  improved decision-making. For more effectively optimized analytics, several IoT data-specification characteristics should be considered. These include dealing with huge volumes of data streamed from sensor-based devices deployed for IoT applications and services. It requires new means of big data analytics that can deal with huge volumes of sensor-generated data.  In this context, conventional hardware/software methods for storage, data analytics and  management purposes cannot handle such huge volumes of streaming data . Moreover, information collected from heterogeneous devices result in three significant common features among IoT Data. These features include data heterogeneity and association of time/space stamps based on the sensors’ locations. The third feature is the subjectivity of IoT data  associated with the high noise levels during acquisition and transmission processes.

\par
Beyond such characteristics that utilize big data analytics approaches, a new suite of applications and services arise that demand prompt actions in real-time analytics. This is primarily due to its time sensitive and fast streaming of IoT data that is generated within short time intervals for instant decision making and actions. These insightful decisions are time stringent, where IoT streaming data analytics need to be delivered within a range of hundreds of milliseconds to only a few seconds. As such, life-saving applications demand fast and continuous streams of incoming data associated in some cases with real-time multi-modal data sources for efficient decision making. For instance, connected and autonomous vehicles require data fusion of real-time sensor data from different sources (e.g., Lidars and cameras), V2X communication and road entities (e.g., traffic lights) for safe perception decision making. Transmitting traffic data to the cloud servers for real-time analytics will be liable to  network and communication latency that are not well-suited for time sensitive applications, which may result in fatal traffic accidents. However, analyzing real-time streaming data on powerful cloud computing platforms that adopt data parallelism and incremental processing techniques can reduce the end-to-end delay associated with two-way data transmissions. A more optimized approach could reside in solutions such as edge computing, where data analytics are closer to the data sources (e.g., edge or IoT-based devices) for faster data analytics \cite{RN201}. However, these solutions are still prone to a number of  limitations including  limited computation, power and storage resources on IoT devices. The rising trends towards real-time data analytics are also striving in non-critical business applications.

\subsection{Decision Making}

Leading IoT-based business sectors rely heavily on well-analyzed real-time data inferred from their IoT-enabled products. For critical and unbiased decision making, real-time data analysis by machine learning algorithms can assist in eliminating/reducing junk information  and estimating learning useful patterns. Data-driven analytics will provide more in depth insights for optimizing customers’ experiences through daily behaviors and patterns analysis. As an example, Apple watches can monitor our daily exercises and sleeping patterns in real-time and assist in providing customized preference notifications. Uber can also make informed decisions based on analyzing real-time demands for traffic trips. This determines their pricing rates that proportionately increase in rush hours. Other examples may include placing sensors within oil tanks for real-time monitoring of oil fluid levels, temperature and humidity. This initiates automated decision making such as oil reordering and planning pre-scheduled maintenance \cite{RN103}. 
 
\par
Decision making-based systems can be classified according to the different levels of system complexity. This includes visual analytics systems that help  business practitioners  to analyze and interpret gathered IoT data. Business intelligence embedded dashboards aid in  presenting the retrieved IoT information in a meaningful manner. Automated and warnings-based systems conduct a predefined data analysis that assists in highlighting risky situations through alerts and warnings. For example, IoT-based real-time environmental monitoring systems can track pollutants and chemicals’ levels in the air within an industrial city. Warning notifications are then subsequently sent to citizens within the affected geofenced area indicating health risk hazards. Reactive-based systems may take a step forward towards performing actions described through rule-based languages that are carried out when specific conditions are met. For instance, smart lighting IoT-based systems may switch off the lights in a specific building area if no one is present, which is indicated by infrared occupancy sensors \cite{RN101}. 

\subsection{Predictive Maintenance}

Utilizing IoT applications has incredibly reduced maintenance costs, in particular in the industrial sector.  For example, industrial equipment manufacturing that embed sensors in heavy machinery integrate with analytical tools to monitor the operational efficacy, detect faults or failures, and provide a full assessment of the operating condition \cite{RN112}. This comprehensive performance evaluation occurs on a regular basis to maintain the system's efficiency and initiates maintenance if needed. This procedure is known as Predictive Maintenance (PdM), or condition-based maintenance, and it employs diagnostics and prognostics data to spot early signs of failure, allowing the system to operate as intended \cite{RN113}. Furthermore, PdM can estimate degradation of the equipment and predict the remaining useful life (RUL) of equipment,  which reduces the maintenance costs to the minimum and assures service availability. According to Selcuk \cite{RN110}, IoT-based predictive maintenance increases the return on investment by 10 times, where this approach increases the total production by 15\%–70\% and reduces the maintenance costs by 25\%–30\%.  Although PdM successfully reduces the cost of production and maintenance, it is expensive to implement due to the high cost of the hardware and software required to effectively incorporate the PdM into the system. Moreover, the quality of the training services and the amount of data required to ensure the efficacy of the PdM performance can be challenging \cite{RN111}.

\subsection{Operation Optimization and Automation}

Industry 4.0 is transitioning from a concept-based approach into a market reality. Through the integration of intelligent and computerized robotic devices into many aspects of industry verticals (e.g, 3D printing, E-sports) that can  assist in automating and optimizing the  manufacturing operations. This allows accurate, timely and cost effective completed manufacturing processes among a set of machines with minimal or no human interventions.  In addition to decrease in cost-related operations through effective inventory management and energy consumption optimization. Effective inventory  management in logistics and supply chain sectors is obtained through the integration of IoT with Radio Frequency Identification (RFID)  \cite{RN102} and barcode scanners. 

\par
Furthermore, IoT technologies within business automation can efficiently be used  for controlling and monitoring machines’ manufacturing operations, performance and rate of productivity through internet connectivity. Moreover, real-time analysis generated from onsite IoT-based sensors provides valuable insights to initiate more efficient ways for decreasing cost-related operational expenses and safety-related/unplanned maintenance issues.  For example, incases of machine operational failures, an IoT-based system can promptly send a machine repair request to the maintenance department for handling the issue.  Furthermore, with the introduction of IoT technologies, business revenues are subsequently expected to increase due to the incremental rise in operational productivity. This can be achieved through  analyzing three critical aspects including operational data, timing and the reasons for any production issues. This allows business leaders to be focused on their high-level core business objectives with a well-defined automated workflow.

\begin{figure*}[!t]
\centering
\includegraphics [width = 7.0in] {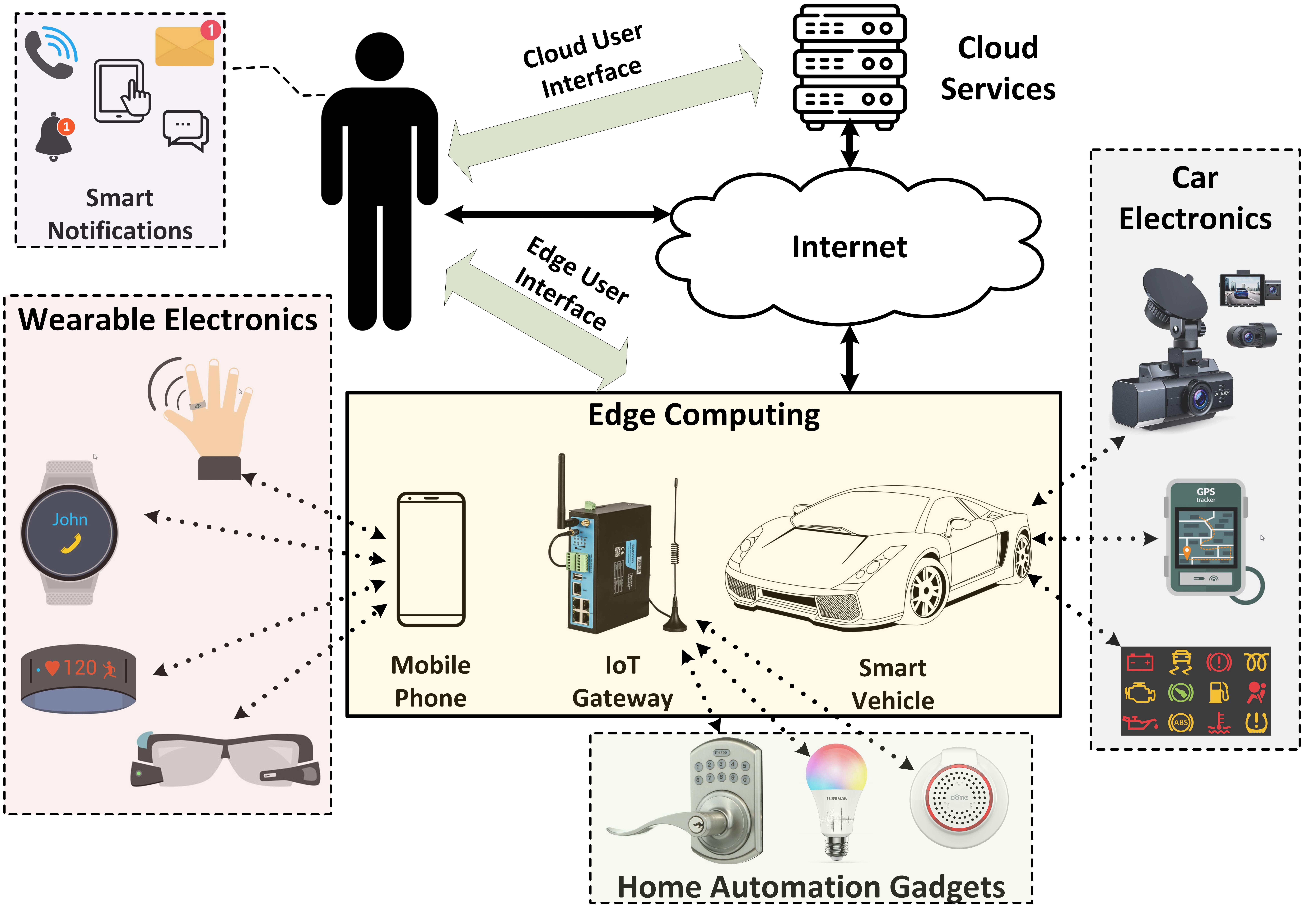}
\caption{User-centric IoT application scenarios.}
\label{figure:6}
\end{figure*}

\subsection{Enhanced Customer Experience}

Connected environments enable businesses to adopt a user-centric approach which utilizes IoT technologies for enhanced overall customer experience and extend the customers’ loyalty towards their services and products. IoT-driven businesses are the gateway to futuristic enhanced digital customer experience and prolonged loyalty which are considered one of the primary laser-focus objectives by many brands. The drive towards more personalized services and applications by customers urge many enterprises to increase services for customer engagement through the use of artificial intelligence-based customer support systems for real-time assistance.

\par
The aforementioned notion of providing level up services that incorporate personalized experiences initiated many innovative applications and services. As such, omnichannel customers’ applications and products such as smart-based home appliances and devices including Alexa-supported devices, Nest Thermostat and intelligent Ring Doorbell cameras that enable customers to use voice assisted technologies along with IoT to control various aspects of their home intelligently. Moreover, ubiquitous smart wearable devices such as fitness trackers that collect real-time health data related to customer behavior and daily routines to enhance customer experience. For instance, providing customers with real-time personalized notifications according to their daily activities. Figure \ref{figure:6} demonstrates the users-centric experience among various IoT services and applications.  

\par
However, privacy and security data leakage are still considered a major challenge that many researchers and developers are trying to find innovative and tangible solutions to secure personal information when shared for improved service and application experiences.

\subsection{Asset Tracking and Monitoring}

Introducing artificial intelligence (AI) into IoT applications has created significant opportunities for innovations in automation and asset tracking domains. Companies and labour-intensive corporations are investing in autonomous working environments with less human interaction, and the demand for AI and context-aware systems has drastically increased. Whereas in times similar to the coronavirus pandemic, factories and workplaces have entirely shut down because of lockdown measures to prevent human interactions. The fusion between AI and IoT transitioned traditional industry models to the industry 5.0 revolution. AI and IoT as core technologies to industry 5.0 along with wireless sensor networks result in more benefits to industries like using analytical techniques to provide predictive maintenance notifications directly affecting downtime, improving workforce and increasing production efficiency.
IoT sensors and devices can perceive and sense their environment through high-level technologies, such as laser scanners, cameras and image processing, movement and proximity. Therefore, real-time decisions are made autonomously regarding object identification and asset tracking by coupling these technologies (i.e., image-recognition software). Similarly, IoT applications based on AI algorithms can learn and think logically about different operations that require problem-solving schemes. Autonomous applications operate based on the receding-strategy approach, where new and old control inputs are carried out simultaneously through computing the new control inputs and executing the old ones. The application creates these control inputs to provide real-time performance based on three hierarchical levels. The higher level is concerned with defining complex operations, for example, GPS waypoints for an autonomous device (e.g., robot) to follow. The other two (mid and lower) levels are precisely related to creating and tracking a reference trajectory for this course, respectively. The safety of the robot is the responsibility of the mid-level controller \cite{RN106}. More concisely, safety concerning collision avoidance among a group of automated robots performing a specific task can be avoided by sharing their perceived data. Other communications that involve Human-Robotic communication are based on the models, such as imitative learning and artificial neural networks.

\section{Open Challenges}
\label{Open_Challenges}

\subsection{Device and Data Heterogeneity}

The versatility of IoT devices and sensor nodes in various fields has given rise to many applications. While deep learning, AI, and many other enabling technologies assist IoT devices in learning by experience and adapting to new environmental inputs to be able to conduct complex operations. However, the reliance on receiving data representing the context of the environment surroundings specific to the IoT application requires a significant number of different sensors and devices. Individual sensors or applications provide limited cognition and visibility of the surrounding environments. Therefore, integrating various sensors is essential in context-aware applications. Furthermore, the diversification of sensor nodes and devices raised numerous challenges in the research community and the industry with respect to unification and standardization. In public sensing, different types of sensors are used (e.g., RFID, Ultrasonic, Cameras, Lidars, etc.) to solve designated issues like real-time counting of people waiting to be served at a specific service provider. The same extends to IoT applications similar to the public sensing domain like traffic management and predictions. Therefore, the demand for modular platforms with unified application programming interfaces (APIs), transmission protocols, data transformation and storage is growing. Moreover, data conversion and normalization operations carried out in applications with heterogeneous devices increase exponentially due to the diversified number of sensors utilized just for one application (e.g., autonomous vehicles).

\par
Behmann \cite{RN109}  described current IoT solutions as point solutions where they are isolated and cannot interact with each other. Collaborative IoT (C-IoT) \cite{RN109} is a recent trend that is still unsaturated and needs more effort to be deployed in real-world scenarios. Sharing the infrastructure and data becomes inevitable to pave the way for C-IoT systems. C-IoT can create an expandable ecosystem, and the IoT community will solve complicated problems by relying on the collaboration between IoT systems. For instance, an ambulance in the emergency state can always have a green light on its way if the emergency service can share the ambulance's route with the city's intelligent traffic system using the shared infrastructure of C-IoT. Recently, an active movement to have a unified standard in different IoT layers has been raised for a few years to mitigate incompatibility challenges faced by the C-IoT trend.

\subsection{Interoperability}

The diversity of IoT devices in the perception layer brings the flexibility to build customized IoT solutions and cherrypick the appropriate device for a specific task that matches constraints regarding the accuracy, cost, compatibility with the existing infrastructure, etc. However, this also comes at the cost of the absence of a unified ecosystem for IoT. This results in interoperability issues between different IoT systems and increased development time to get diverse IoT devices to act as a coherent system. Besides, it leads to the notorious ``isolated islands" of miniature IoT subsystems based on the brand of devices or their enabling technologies, hindering the utilization of sensed information by different systems to its maximum and impeding the potential promise of IoT.

\subsection{User and Data Privacy}

The constituent IoT gadgets of an IoT system typically consist of consumer electronics (e.g., smart TVs) and wearable devices (e.g., smart watches) that gather a lot of information about people, which was previously hard to collect. Gathered data by IoT devices may include personal information of the users (e.g., name, birthdate, etc.), their biometric information (e.g., fingerprint, voice recognition, etc.), and their preferences (e.g., eating habits, preferred movie genres, etc.), which are usually part of the device's initial setup, registration to its cloud platform, or for the device to be able to perform its smart designated task efficiently. Moreover, advanced IoT systems typically involve the aggregation of numerous pieces of information from heterogeneous smart objects, which is known as ``sensor fusion" \cite{RN128} to provide accurate and comprehensive data about the environment, including people themselves, to help make better informed decisions. Thanks to the advancements in artificial intelligence (AI) domain technologies which can leverage granular data collected by smart objects to generate inferences that would not be achievable with coarser data from individual smart objects. This intelligence imparted to IoT catalyzes its wide adaption and makes it quite useful in different application domains.
However, user privacy concerns are still an open challenge to IoT that impedes its widespread adoption and limits its potential promise. IoT systems can disclose identifiable information about people without their consent. Therefore, amidst the potential promise of IoT to change the way we deal with our surroundings, users are mostly worried about the potential of private information leakage \cite{RN129}. They are worried about who owns their data and how it is utilized. Nonetheless, the notorious correlation between service providers and device vendors on one hand, and data brokers on the other hand, raises concerns about the possibility of their personal information being disclosed for non-public interest objectives. People frequently alter their behavior when they suspect that their identifiable information and activity footprints are being monitored, which reduces their freedom, changes their lifestyle, and makes them sceptical of IoT. In the following subsections, we review data privacy concerns that are associated with IoT.

\subsubsection{De-identification of IoT Data}

Generally, it is prohibited to make datasets that include identifiable personal information publicly accessible. One common way to prevent exposing personal information in datasets is to avoid gathering information that could be used to identify people in the first place and whenever possible. For example, PIR sensors could be used for occupancy detection rather than surveillance cameras. However, given the penetration of IoT in a lot of domains with differentiated requirements, it is usually hard, or even impossible, to preclude the inclusion of identifiable information in gathered datasets by IoT. In this context, de-identification \cite{RN130} is the process used to anonymize identifiable personal information from datasets, which is quite challenging. Hashing algorithms are commonly used to pseudonymize identifiable information in datasets by replacing identified people in a dataset with their unique hash token. However, since different datasets often have a lot in common, it is usually easy to figure out who the hashed information belongs to using inference techniques.

\subsubsection{Consent}

Consent \cite{RN131} is the typical justification for businesses to collect, use, and disclose personal information. However, consent often matters more than just unconsciously clicking the ``I agree" button by the end user on the ``Terms and Conditions" statement page of a device. Rather, consent that is meaningful and effective requires well-defined and finely-grained structured objectives that the user should be able to choose from. Moreover, one can not presume that consent will last forever. Therefore, consent methods should represent a single acceptance at a single moment in time, which may not be suitable for the continual nature of IoT. Also, given the interoperability nature of the IoT, where a smart sensor node may be utilised by different systems with different privacy policies, an individual can not grant meaningful permission for the use of their personal data for vague or broad purposes.

\subsection{Vendor Lock-in}

IoT vendors and service providers typically maintain the security of their active devices or services by regularly providing firmware patches and system updates that address security vulnerability issues that continually emerge. However, they often have different expectations about how long their products or services will last than the people who buy them. For instance, vendors may terminate technical support or firmware maintainability of a device, or the service provider may discontinue the service that the device relies on to operate, far before the end user plans to retire the device. This usually comes at the cost of possible security holes, privacy issues, and vendor lock-in \cite{RN132}. Therefore, customers would have to stick with their active line of products and services to keep their systems safe and operational because suppliers would no longer care about security and privacy issues with their retired devices or have the skills to deal with them. 

\subsection{Device Management}

The ``plug and play" feature that usually accompanies IoT devices makes them user-friendly since customers can seamlessly get them up and running effortlessly without the need for complicated setup procedures. However, this sometimes comes at the cost of potential user privacy exposure since the default setup of devices usually comes with insufficient privacy and security precautions. Nonetheless, the fact that a gadget is an IoT device that can collect personal information and send it to third parties over the cloud may not be even realized by the majority of non-technically savvy customers by default. In addition, most IoT customers find it hard and time-consuming to adjust the privacy settings for each device in the system. This is because IoT does not have a standard ecosystem and is often made up of many devices from different manufacturers, each with their own user experience interface. 

\subsection{Accountability}

The extensive and distributed nature of IoT systems, which typically include different service providers that handle the collected data to achieve the designated task of the system, makes it challenging to precisely determine who is responsible for what. In order to achieve a robust and highly reliable IoT system architecture, system designers usually follow a common system architectural model that is known as ``microservices" \cite{RN133}. They divide the ultimate task of the system into small independent tasks, each of which may utilize numerous services from different service providers that communicate over well-defined APIs. However, this raises privacy concerns because the collected data, which may contain personally identifiable information about users, is now handled and commonly stored by various third-party organizations with hazy boundaries that may apply different privacy policies.

\subsection{Transparency}

The tiny size of most IoT devices in use today without a screen, or at least an adequate screen size to display a lot of textual information, makes it difficult for users to review their privacy policies before they start using them. In order to review the privacy policies of these devices, however, users should login to the website of the manufacturer of the device or use a proprietary software or a mobile application for the device. However, in either case, given the heterogeneous nature of the IoT and the anticipated large number of IoT devices people use on a daily basis, it looks extremely challenging to follow the privacy policy of each encountered device. Also, a lot of privacy policies for IoT devices seem vague to the majority of people. Moreover, some IoT devices that exist in organizations and public settings are usually anonymous without details about the type of information they collect or how this information is utilized and for which objectives. Also, most of the time, users do not have the option to stop the collection.

\subsection{Security (Data and Device Vulnerability)}

IoT devices exchange data with millions of devices through the internet which implicitly exposes the IoT devices to the vulnerabilities and security threats of the Internet protocol stack \cite{RN138}. The amount of data collected, stored and shared between IoT devices and the service providers are expected to grow significantly. Besides the extraordinary amount of data produced by IoT devices, they induce evidently high-security risks and potential cyberattacks destabilizing many applications and industries.

\par
IoT networks come with their unique security challenges \cite{RN139}, where each layer is exposed to certain types of attacks \cite{RN140}, like Distributed Denial of Service attacks (DDoS) on the network layer. To that extent, a multitude of surveys cite{refs2} and studies have been conducted to expose existing security threats and vulnerabilities in current IoT applications. Recent surveys on IoT data and device security emphasize that the gap between applying existing security techniques to emerging IoT applications is growing significantly. The security gaps in IoT applications are categorized into vendor-related security issues and available resources or capabilities on the IoT nodes. IoT vendors for sensors and devices are moving towards low-cost manufacturing that lack security features. Similarly, the heterogeneity of the IoT applications, protocols, and hardware increases the security scope of threats in IoT applications \cite{RN140}. On the other hand, IoT applications are inherently constrained by the limited processing and storage capabilities of devices to carry out sophisticated security techniques. Therefore, new security measures are introduced for IoT resource-constrained devices using robust ML techniques like the TinyML framework. The concepts introduced behind the integration of ML is to increase the flexibility of IoT nodes in defending against emerging security threats \cite{RN141}. IoT devices can then train the deployed ML models to work against new security threats.  

\subsection{Open Deployments and Access Control}

Mentioning the access control usually flashed RFID (Radio-frequency identification) cards. RFID technology sparked the existence of the IoT term coined by Kevin Ashton, who considered RFID a vital component in The IoT systems. Access control as an open challenge is a multifaceted challenge that has been raised due to other issues. In this section, we use a wallet holding many RFID cards as an example to discuss the access control issues:
\begin{itemize}
  \item Heterogeneity: Having many cards to access different purposes itself is due to the heterogeneity of the systems. There are chances to have a consensus among some corporations to unify their access cards to mitigate this challenge. For instance, Google Pay is an android application that offers contactless purchases on the smartphone with a built-in Near Field Communication NFC module. Users can register debit or credit cards and use them in their daily in-person shopping instead of holding many bank cards in wallets. Due to systems heterogeneity, we still face a challenge in making all services accessible from unified access.
  \item Security: Giving accessibility to banks a counting using contactless RFID bank cards looks a very smooth user experience in transactions instead of writing sixteen digits in card readers. However, card skimming devices can clone contactless bank cards.
\end{itemize}

To discuss the other issues in access control, we use a smart home application as another example facing access control challenges:
\begin{itemize}
  \item Interoperability: Recently, users can simply control their IoT devices in their houses, such as smart TV, fridge, coffee machine, and adapted light systems. The interoperability between these devices is still very challenging due to the lack of standardization. For instance, a coffee machine starts pouring coffee on a cup if the adapted light senses a motion in the living room, and there is a collaborative integration between these devices. ThingsDriver \cite{RN114} is a beginning to have A Unified Interoperable messaging protocol that, if adopted by cooperates, can pave the way to have a collaborative environment.
  \item Privacy: All these smart home devices become remotely accessible through user-friendly user interfaces such as a smartwatch or phone. On the other side, the flexibility of accessibility raises privacy concerns. Residents' data are highly vulnerable to being breached by unauthorized access.
\end{itemize}

\section{Conclusions}
\label{Conclusion}

In this paper, we review the Internet of Things technology from different architectural, technological, operational and value-proposition perspectives. We first shed the light on the definition, acclaimed value and potential, and unique features and characteristics compared to similar previous technologies and standard layered architecture. We then highlight the different applications of IoT in various life domains that primarily benefit from its realization as a novel computing paradigm. We outlined the grand challenges facing IoT which may cause slowdown in its widespread adoption at the individual, organizational and governmental levels. 

\par
We believe the IoT will continue to grow as a disruptive technology that changed the world and it will never be the same again. There is a continuously increasing reliance on IoT technology in different sectors of our life for its convenience and innovative applications stemming out of it. Individuals and enterprises started to gain confidence in the technology and overlook or ignore the downsides of its security and privacy aspects. However, we also believe that the emergence of Edge computing in its different forms and shapes has contributed to lower the adoption barriers of IoT and increased interest in its technology and smart services. We anticipate that in the next few years IoT will continue to penetrate deeper in various sectors and tape into more industrial and governmental settings.

\bibliographystyle{IEEEtran} 
\bibliography{iotchallenge}




\end{document}